\documentclass[preprint,amsmath,amssymb]{revtex4}
\newcommand {\pa}{\partial}

\newcommand {\fr}{\frac}
\newcommand {\vphi}{\varphi}

\newcommand {\beg}{\begin{equation}}
\newcommand {\en}{\end{equation}}
\newcommand {\bega}{\begin{eqnarray}}
\newcommand {\ena}{\end{eqnarray}}
\begin{document}

\title{Dark energy from gravitoelectromagnetic inflation?}
\author{ $^{1,2}$ Federico Agust\'{\i}n Membiela\footnote{E-mail address: membiela@argentina.com}
and $^{1,2}$ Mauricio
Bellini \footnote{E-mail address: mbellini@mdp.edu.ar}}

\address{$^{1}$ Departamento de F\'{\i}sica, Facultad de Ciencias Exactas y
Naturales, Universidad Nacional de Mar del Plata, Funes 3350,
(7600) Mar del Plata,
Argentina.\\
$^2$ Consejo Nacional de Investigaciones Cient\'{\i}ficas y
T\'ecnicas (CONICET), Argentina. }

\begin{abstract}
Gravitoectromagnetic Inflation (GI) was introduced to describe in
an unified manner, electromagnetic, gravitatory and inflaton
fields from a 5D vacuum state. On the other hand, the primordial
origin and evolution of dark energy is today unknown. In this
letter we show using GI that the zero modes of some redefined
vector fields $B_i=A_i/a$ produced during inflation, could be the
source of dark energy in the universe.
\end{abstract}
\maketitle

%\pacs{04.20.Jb, 11.10.kk, 98.80.Cq}

\section{Introduction}

It is believed that our universe has experimented a primordial
accelerated expansion called inflation, in which the equation of
state was vacuum dominated: $\left.p/\rho\right|_{{\rm infl}}
\simeq -1$. This theory explains the origin of large scale
structure formation\cite{infl} and could also explain the origin
of primordial magnetic fields\cite{mag}. New approaches to
inflationary cosmology based on a scalar field formalism have been
proposed in the context of higher-dimensional theories of gravity.
In particular using ideas of the induced matter theory\cite{imt} a
new formalism for describing inflation, scalar metric fluctuations
and gravitational waves, has been introduced\cite{waves}. The
basic idea of this formalism is the existence of a 5D space-time
equipped with a purely kinetic scalar field $\varphi$, in which
our observable 4D universe can be confined to a particular
hypersurface by the choice of a particular frame.

The GI theory\cite{mag} was developed with the aim to describe, in
an unified manner, the inflaton, gravitatory and electromagnetic
fields during inflation. Other 4D\cite{ba} and 5D\cite{..}
formalisms were developed in the last decades. This approach has
the advantage that all the 4D sources has a geometrical origin in
the sense that they can be geometrically induced when we take a
foliation on the spacelike and noncompact fifth coordinate. GI was
constructed from a 5D vacuum state on a Riemann-flat metric and
can explains the origin primordial magnetic fields on cosmological
scales, which are today observed. In this letter we explore, using
this theory, the possibility that dark energy can be produced
during inflation. Dark energy plays a very important role in the
today observed accelerated expansion of the universe. However, the
physical explanation of its origin and evolution is still
mysterious.

\section{Formalism}

To develop our formalism we shall consider a 5D space-time
described by the line element
$dS^2=g_{ab}\,dx^a\,dx^b$
\begin{equation}\label{..}
dS^2=e^{2F(\psi)} h_{\mu\nu}(x^{\mu}) dx^{\mu} dx^{\nu} - d\psi^2,
\end{equation}
where $g_{ab}$ is the 5D covariant tensor metric, $h_{\mu\nu}$ is
the 4D covariant tensor metric,  $a,b$ run from $0$ to $4$, Greek
letters run from $0$ to $3$ and latin letters run from $1$ to $3$.
Inflation from a warped space was recently studied in\cite{ed}.

Now we consider the system described by the action
\begin{equation}
^{(5)}{\cal S} = {\Large\int} d^5 \sqrt{|g|} \left[\frac{^{(5)}
{\cal R}}{16 \pi G} + ^{(5) } {\cal L}\left(A_b,
A_{c;d}\right)\right],
\end{equation}
for a vector potential with components $A_a=(A_{\mu},\varphi)$,
which are nonminimally coupled to gravity
\begin{equation}
^{(5)} {\cal L}\left(A_a, A_{a;b}\right)= -\frac{1}{4} Q_{bc}
Q^{bc} - \frac{^{(5)} {\cal R}}{6} A^b A_b,
\end{equation}
such that $^{(5)} {\cal R}$ is the 5D Ricci scalar and
\begin{equation}
Q_{bc} = F_{bc}+ \gamma\,g_{bc}\left(A^b_{\,;b}\right)^2,\qquad
\gamma = \sqrt{(2\lambda/5)}, \qquad F_{bc} =A_{c;b} -A_{b;c},
\end{equation}
where $(;)$ denotes the covariant derivative. The equations of
motion for the vector components $A^b$ are
\begin{equation}\label{motion}
\left(\triangle A\right)^{b} -{\cal R}^{b}_{\,\,c} A^c+(1-\lambda)
A^{c\,\,\,\,\,;b}_{\,\,;c} =0,
\end{equation}
where $\left(\triangle A\right)^{b}=-A^{b;c}_{\,\,\,;c}
+R^{b}_{\,\,c} A^{c}$ denotes the Rham vector wave operator, which
is a generalized d'Alambertian for vectors in curved spacetime. We
shall consider the Lorentz gauge: $A^c_{\,\,;c}=0$. Notice that
the $\alpha$ (Greek letters run from $0$ to $3$) components of
(\ref{motion}) with the Lorentz gauge on a Ricci flat metric
(${\cal R}^a_{\,\,b}=0$), give us the 5D Maxwell equations without
sources: $A^{b;c}_{\,\,\,;c}=0$. One could obtain Maxwell's
equations with sources on an effective 4D space-time by taking a
foliation on the fifth coordinate of this 5D flat metric in the
particular gauge $A_4=0$. In this case the effective
electromagnetic current has only 4 non-zero components which are
purely of 5D origin.

\section{5D kinetic vector field and effective 4D vector and scalar fields}

We wish to study a period of inflation, where there exist scalars
and vector fields that are affected by an extra dimensional vector
field\cite{mag}. During inflation the field that dominates is a
neutral scalar field called inflaton. It was shown that also a
collection of vector fields may produce also a period of
inflation\cite{mukha}. Our study tries to give an unified
treatment for both formulations. In principle, we may obtain
scalar and vector equations of motion for all of the Kaluza-Klein
modes, but only a window of these modes can survive a period of
inflation. These remanent modes (more exactly, the zero modes)
would carry the energy of the universe that after the end of
inflation decay to fundamental particles. On the other hand,
finite $k$-modes should be responsible for the seeds of magnetic
fields and structure formation in the universe once the
inflationary period ends. For simplicity, in this letter these
modes will be neglected. Only we shall deal only with the zero
$k$-modes of the components $A_c$, of the vector potential.

We can define the vector $B_i(t,\vec r,\psi)\equiv A_i(t,\vec
r,\psi)/a(t)$. We work with a Friedmann-Robertson-Walker (FRW) 4D
metric $h_{\mu\nu}=diag[1,-a(t)^2,-a(t)^2,-a(t)^2]$. We are
considering the 5D Lorentz gauge, but in the absence of sources we
can still use the radiation gauge to eliminate the $A_0$ equation.
In this case the 5D Lorentz gauge reduces to

 \beg
    e^{2F}\left(4F'+\fr{\pa}{\pa\psi}\right)A_4+a^{-2}\vec{\pa}\cdot\vec{A}=0,
 \en
where $\vec{\pa}\cdot\vec{A}\equiv \left(A_i\right)^{;i}=0$
denotes the 3D divergence of $\vec A$.

\subsection{5D invariant $I$}

Taking into account that $I= A^b \, A_b$ is a 5D scalar, which is
invariant under coordinates transformations and characterizes the
strength of the vector field on 5D. In particular, on the metric
(\ref{..}) it takes the explicit form
\begin{equation}\label{expl}
I=e^{-2F} \left[A^2_0 - A^i\,A_i\right] - \varphi^2,
\end{equation}
such that its variation is null: $\delta\,I=0$. Taking $A_0=0$,
one obtains
\begin{equation}
I = -\left[e^{-2F}  A^i\,A_i + \varphi^2\right].
\end{equation}
In particular, for 5D Riemann flat metrics the scalar $I$ in
(\ref{expl}) becomes a constant.

\subsection{Spatially homogeneous dynamics of fields}

Taking null spatial derivatives we work with homogeneous fields
which are given by their 3D expectation values. In this case the
gauge reduces to
 \beg
    \fr{\pa A_4}{\pa\psi}=4F'A_4,
 \en
and the equations for the vector and scalar reduce to
 \bega
    \left<\ddot B_i\right>+3H\left<\dot B_i\right>-\left[2(4F'^2+F'')e^{2F}-(2H^2+\dot H)-e^{2F}
    \left(\fr{\pa^2}{\pa\psi^2}+2F'\fr{\pa}{\pa\psi}\right)\right]\left<B_i\right>=0, \label{h1}\\
    \left<\ddot A_4\right>+3H\left<\dot
    A_4\right>-e^{2F}\left[\fr{\pa^2}{\pa\psi^2}+4(6F'^2+F'')\right]\left<A_4\right>=0, \label{h2}
 \ena
where $\left<B_i\right>$ and $\left<A_4\right>$ give us the
$(t,\psi)$-dependent expectation values calculated on the 3D
spatial volume
 \bega
    \left<B_i\right>&=&B_i(t,\psi),\\
    \left<A_4\right>&=&\vphi(t,\psi).
 \ena
In general, the difference between the scalar and vector fields is
that (despite their expectation values may be spatially
homogeneous), the last ones are essentially anisotropic. Hence,
they can serve to explain the existence of primordial cosmological
magnetic fields on sub Hubble scales. From another point of view,
they can also contribute to an important amount to the energy
density of the universe during, but mainly, after inflation. Once
we assume the isotropy of $\left<B_i\right> \equiv B$, we note
that the equation of motion (\ref{h1}) for $B$, has the same
mathematical structure than whole of the 3D expectation value for
the inflaton field $\left<A_4\right>=\varphi(t,\psi)$.

\section{An example with the Ponce de Le\'on metric}

As an example we can consider the case where $F(\psi)={\rm
ln}(\psi/\psi_0)$, with $a(t)\sim e^{t/\psi0}$. This case is very
interesting because we obtain the Ponce de Le\'on 5D Riemann flat
metric
\begin{equation}\label{pdl}
dS^2= \left(\frac{\psi}{\psi_0}\right)^2 \left[dt^2 -
e^{2t/\psi_0} d\vec{r}^2\right] - d\psi^2,
\end{equation}
which is 3D spatially flat, isotropic and homogeneous. This means
that any invariant defined on the Riemann flat 5D metric
(\ref{pdl}) will be a constant. If we take a foliation
$\psi=\psi_0=a/\dot a =1/H$, we obtain the effective 4D de Sitter
metric
\begin{equation}\label{4d}
\left. dS^2\right|_{eff} =dt^2 - e^{2 H\,t} d\vec{r}^2,
\end{equation} with an effective geometrically induced 4D scalar
curvature $^{(4)} {\cal R}= 12 H^2$ (In this particular case the
Hubble parameter is a constant of time)\footnote{Here, we have
used ideas of space-time-matter theory to induce the scalar
curvature of the effective 4D hypersurface: $^{(4)} {\cal R}=
\left.12/\psi^2\right|_{\psi=1/H}=12 H^2$. See for instance, P. S.
Wesson. {\em Space-Time-Matter: Modern Kaluza-Klein Theory (World
Scientific: Singapore, 1999).}}. Furthermore, the scalar $I$ is a
constant on the Ponce de Le\'on metric and once we take the
particular foliation $\psi=1/H$, on comoving coordinates
$U^x=U^y=U^z=0$, one obtains
\begin{eqnarray}
\left.I\right|_{\psi=1/H} &=& \left. H^2
\left<A_i\right>\,\left<A^i\right>+\left<\varphi\right>^2\right|_{\psi=1/H}\nonumber
\\
&=& \left. H^2
\left<B_i\right>\,\left<B_i\right>+\left<\varphi\right>^2\right|_{\psi=1/H}={\rm
const},
\end{eqnarray}
where we assumed the summation over repeated spatial indices. It
is well known that when slow rolling conditions are assumed
$\left<\varphi\right>(t)$ is a constant of time for inflationary
models which describe a de Sitter (exponential)
expansion\footnote{The reader can see it after the eq. (14) of the
paper: A. Membiela and M. Bellini, Phys. Lett. {\bf B635}, 243
(2006).}. Hence, during inflation $\left<B_i\right> $ should be a
constant and $\left<A_i\right> \sim a$. Notice that {\em
$\left<B_i\right>$ has nothing to do with magnetic fields}. These
field components are re-scaled zero-modes of $A_i$.

\section{Final Remmarks and possible generalizations}

Notice that in the example here developed we have worked an
effective de Sitter expansion which becomes after taking a
constant foliation $\psi=1/H$ on the fifth coordinate. However,
one could obtain more general inflationary models where the scale
factor grows quasi-exponentially by using a dynamical foliation on
the fifth coordinate $\psi \equiv \psi(t)$. In that case one
obtains the more general condition
\begin{equation}
\left.\frac{d}{dt} \left( H^2 \left< B_i\right>\,\left<
B_i\right>\right)\right|_{\psi(t)} =\left. -\frac{d}{dt}
\left<\varphi\right>^2\right|_{\psi(t)}
> 0.\label{ul}
\end{equation}
In other words, the term $H^2 \left< B_i\right>\,\left<
B_i\right>$ in (\ref{ul}) grows as the vacuum energy density, and
the term $\left.\left<\varphi\right>^2\right|_{\psi(t)}$, decays.
Hence, the increase of $H^2 \left< B_i\right>\,\left< B_i\right>$
during inflation could be the source of the present day domination
of dark energy in the universe. Notice that this term has an
electromagnetic origin, but it should be very important on very
large (cosmological) scales during inflation.

\end{document}